\documentstyle[prb,aps,multicol,epsf]{revtex}

\begin{document}
\title{Compressible Anisotropic States around the Half-Filled Landau Levels}
\author{Nobuki Maeda}
\address{
Department of Physics, Hokkaido University, Sapporo 060-0810, Japan}
\date{\today}
\maketitle

\begin{abstract}
Using the von Neumann lattice formalism, we study 
compressible anisotropic states around the half-filled Landau levels 
in the quantum Hall system. 
In these states the unidirectional charge density wave (UCDW) state 
seems to be the most plausible state. 
The charge density profile and Hartree-Fock energy of the UCDW are 
calculated self-consistently. 
The wave length dependence of the energy for the UCDW is also obtained 
numerically. 
We show that the UCDW is regarded as a collection of the one-dimensional 
lattice Fermi-gas systems which extend to the uniform direction. 
The kinetic energy of the gas system is generated dynamically from 
the Coulomb interaction. 
\end{abstract}
\draft
\begin{multicols}{2}

\section{Introduction}

Two-dimensional electron systems in a strong magnetic 
field have been providing various fascinating phenomena for the last 
two decades of this twentieth century. 
The integer quantum Hall effect (IQHE)\cite{a} and fractional quantum 
Hall effect (FQHE)\cite{b} are observed around the integral filling factor 
and rational filling factor with an odd denominator\cite{bb} respectively. 
These effects are caused by the formation of the incompressible liquid 
state with a finite energy gap. 
The IQHE state has a Landau level's energy gap or Zeeman's energy gap and 
FQHE state has an energy gap due to the Coulomb interaction.\cite{c} 
Remarkable progress of the composite fermion (CF) theory\cite{d,e} for the 
FQHE shed light on the Fermi-liquid-like state at the half-filled 
lowest Landau level. 
Evidences for the Fermi-liquid-like state were obtained in many 
experiments.\cite{f} 
It is believed that the state is compressible isotropic state and 
has a circular Fermi surface. 

Recent experiments at the half-filled Landau levels have 
revealed the anisotropic nature of the compressible states. 
At the half-filled third and higher Landau levels, 
highly anisotropic behavior was observed in the 
magnetoresistance.\cite{j,k} 
At the half-filled lowest and second Landau levels, transition to the 
anisotropic state was also 
observed in the presence of the periodic potential\cite{l} or in-plane 
magnetic field\cite{m,n} respectively. 
Although the origin of the anisotropy is still unknown, 
the unidirectional charge density wave (UCDW) state is a 
candidate for the anisotropic state. 
Theoretical works in higher Landau levels showed possibility of the 
UCDW.\cite{o,p} 
Recent theoretical works in lower Landau levels support the UCDW or 
liquid-crystal state.\cite{pa,paa,pb} 
In this paper we investigate the compressible charge density wave (CCDW) 
states, which include the UCDW state and compressible Wigner crystal 
(CWC) state, in the several lower Landau levels.  
As a result, the UCDW states are found to be the lowest energy states 
in the CCDW states at the half-filled Landau levels. 

The CCDW state is a gapless state with an anisotropic Fermi surface and 
has a periodically modulated charge density. 
Using the von Neumann lattice\cite{q,t} representation, 
we construct the CCDW state and calculate the charge density profile 
and Hartree-Fock energy self-consistently. 
The von Neumann lattice representation has a quite useful property 
in studying the quantum Hall system with the periodic potential.\cite{u,v} 
The lattice structure of the von Neumann lattice can be adjusted to the 
periodic potential by varying the modular parameter of the unit cell. 
In the present case the periodic potential is caused by the charge density 
modulation through the Coulomb interaction.\cite{va,w,x} 
If the translational invariance on the lattice is unbroken, 
a Fermi surface is formed. 

In the Hartree-Fock approximation, we show that 
the self-consistency equation for the 
CCDW has two types of solution at the half-filling. 
The one has a belt-shaped Fermi-sea corresponding to the UCDW and 
the other has a diamond-shaped Fermi-sea corresponding to the CWC. 
For the belt-shaped Fermi sea, one-direction of the momentum space is filled 
and the other direction is partially filled. 
Therefore the UCDW is regarded as a collection of the one-dimensional 
lattice Fermi-gas systems which was called the quantum Hall gas (QHG) in 
Ref.~\cite{x}. 
In the UCDW state, there are two length scales, 
the wave length of the UCDW, $\lambda_{\rm CDW}$ and Fermi wave length of 
the lattice fermions, $\lambda_{\rm F}$. 
These two parameters obey a duality relation. 
We obtain the wave length dependence of the energy. 
Moreover we calculate the kinetic energy of the gas system. 

The paper is organized as follows. 
In Sec.~II, 
the Hartree-Fock energy for the CCDW states is calculated. 
The density profile, wave length dependence of the energy, and the kinetic 
energy of the UCDW are obtained in Sec.~III. 
The summary and discussion are given in Sec.~IV. 

\section{Hartree-Fock energy for the CCDW state : 
UCDW vs. CWC}

In this section we construct the CCDW state in the Hartree-Fock 
approximation using the formalism developed in Ref.~\cite{x}. 
Let us consider the two-dimensional 
electron system in a perpendicular magnetic field $B$ in the 
absence of impurities. 
The electrons interact with each other through the Coulomb potential 
$V({\bf r})=q^2/{\rm r}$. 
In this paper we ignore the spin degree of freedom and 
use the natural unit of $\hbar=c=1$. 
In the $l$-th Landau level space, the free kinetic energy is quenched 
as $\omega_c(l+1/2)$, $l=0,1,2\dots$, 
where $\omega_c=eB/m$. 

\begin{figure}[h]
\centerline{
\epsfysize=1.8in\epsffile{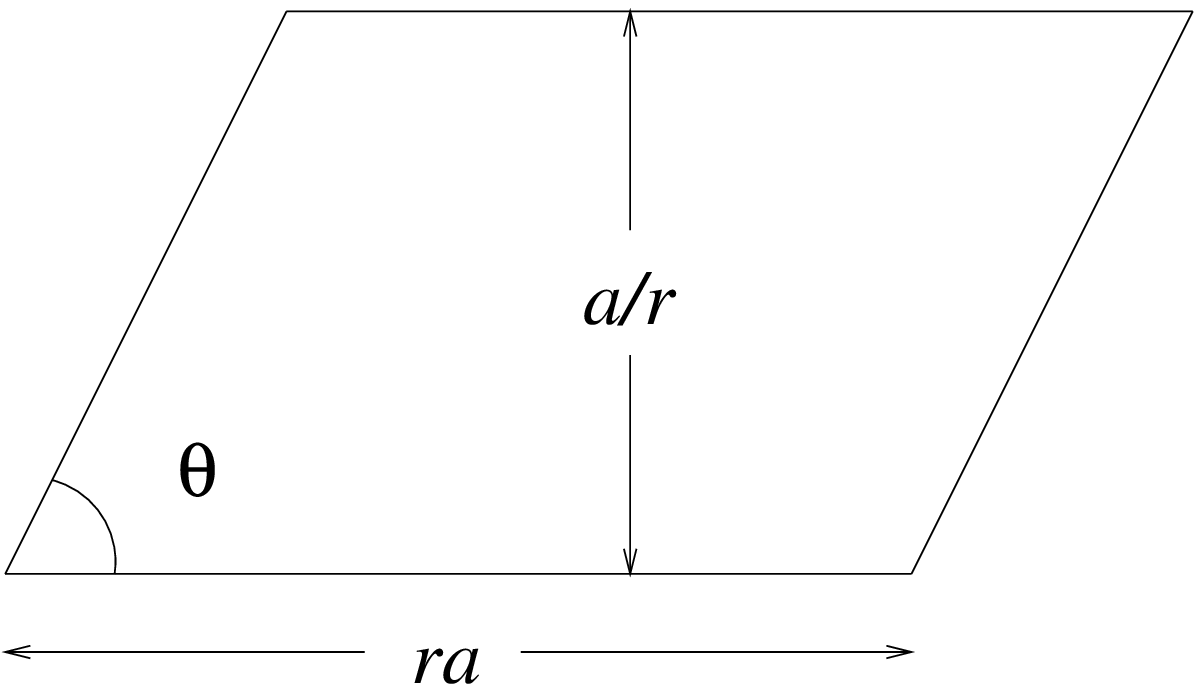}}
FIG. 1 The unit cell of the von Neumann lattice spanned by the 
vectors ${\bf e}_1$ and ${\bf e}_2$.
\end{figure}
In the von Neumann lattice formalism,\cite{t,u,v} 
the electron field is expanded as
\begin{equation}
\psi({\bf r})=\sum_{l,{\bf X}}b_l({\bf X})W_{l,{\bf X}}({\bf r}),
\end{equation}
where $b$ is an anti-commuting annihilation operator and $\bf X$ is 
an integer valued two-dimensional coordinate. 
The Wannier base functions $W_{l,{\bf X}}({\bf r})$ are orthonormal 
complete basis in the $l$-th Landau level.\cite{v}. 
Expectation values of the position of $W_{l,{\bf X}}({\bf r})$ are
located at two-dimensional lattice sites $m{\bf e}_1+n{\bf e}_2$ for 
${\bf X}=(m,n)$, where ${\bf e}_1=(ra,0)$, ${\bf e}_2=(a/r\tan\theta,
a/r)$, and $a=\sqrt{2\pi/e B}$. 
The area of the unit cell is ${\bf e}_1\times{\bf e}_2=a^2$ which 
means that a unit flux penetrates the unit cell of the von Neumann lattice. 
The unit cell is illustrated in Fig.~1. 
For simplicity we set $a=1$ in the following calculation. 

The Bloch wave basis, which is given by $u_{l,{\bf p}}({\bf r})=
\sum_{\bf X}W_{l,{\bf X}}({\bf r})e^{i{\bf p}\cdot{\bf X}}$, is 
another useful basis on the von Neumann lattice.\cite{v} 
The lattice momentum $\bf p$ is defined in the Brillouin zone (BZ), 
$\vert p_i\vert\leq \pi$. 
The wave function $u_{l,{\bf p}}({\bf r})$ extends all over the plane and 
its probability density has the same periodicity as the von Neumann lattice. 

In the momentum space, the system has a translational invariance\cite{x} 
which is referred to as the K-invariance in the CF model.\cite{HS}
In the following, we show that symmetry breaking of the 
K-invariance generates a kinetic energy and leads to an anisotropy 
in the charge density. 

We consider a mean field state of filling factor $\nu=l+\bar\nu$ 
where $l$ is an integer and $0<\bar\nu<1$. 
Let us consider a mean field for the CCDW which has the translational 
invariance on the von Neumann lattice as
\begin{equation}
U_l({\bf X}-{\bf X}')=\langle b_l^\dagger({\bf X}')b_l({\bf X})\rangle,
\label{U}
\end{equation}
and $U_l(0)=\bar\nu$. 
Ignoring the Landau level's energy and 
the inter-Landau level effect, the Hartree-Fock Hamiltonian within 
the $l$-th Landau level $H ^{(l)}_{\rm HF}$ reads
\begin{eqnarray}
H ^{(l)}_{\rm HF}&=&\sum_{{\bf X},{\bf X}'} U_l({\bf X}-{\bf X}')\{
{\tilde v}_l (2\pi ({\hat{\bf X}}-{\hat{\bf X}}'))-v_l({\hat{\bf X}}-
{\hat{\bf X}}')\}
\nonumber\\
&&\phantom{\sum_{{\bf X},{\bf X}'}}
\times\{b^\dagger_l({\bf X}) b_l ({\bf X}')-
{1\over2} U_l({\bf X}'-{\bf X})\},
\label{hart}
\end{eqnarray}
where
\begin{eqnarray}
{\tilde v}_l ({\bf k})&=&\{L_l({k^2 \over 4\pi})\}^2e^{-{k^2 \over 4\pi}} 
{\tilde V}({\bf k}),\\
v_l ({\bf X})&=&\int{d^2k\over (2\pi)^2}{\tilde v}_l ({\bf k}) e^{i{\bf k}
\cdot{\bf X}}.
\end{eqnarray}
Here ${\tilde V}({\bf k})=2\pi q^2/k$ for ${\bf k}\neq0$ and 
${\tilde V}(0)=0$ due to the charge neutrality condition. 
$\hat{\bf X}$ is a position of the Wannier basis in the real space, 
that is, $\hat{\bf X}=(rm+n/r\tan\theta,n/r)$ for ${\bf X}=(m,n)$. 

By Fourier transforming Eq.~(\ref{hart}), we obtain the self-consistency 
equations for the kinetic energy $\varepsilon_l$ as 
\begin{eqnarray}
\varepsilon_l({\bf p},\bar\nu)&=&\int_{\rm BZ}{d^2 p'\over (2\pi)^2} 
{\tilde v}^{\rm HF}_l({\bf p}'-{\bf p})\theta(\mu_l-
\varepsilon_l({\bf p}',\bar\nu)),
\label{self1}\\
\bar\nu&=&\int_{\rm BZ}{d^2 p\over (2\pi)^2}
\theta(\mu_l-\varepsilon_l({\bf p},\bar\nu)),
\label{self2}
\end{eqnarray}
where $\mu_l$ is the chemical potential and ${\tilde v}^{\rm HF}_l$ 
is defined by
\begin{equation}
{\tilde v}^{\rm HF}_l({\bf p})=\sum_{\bf X}\{
{\tilde v}_l (2\pi ({\hat{\bf X}}))-v_l({\hat{\bf X}})\}e^{-i{\bf p}
\cdot{\bf X}}.
\end{equation}
Equations (\ref{self1}) and (\ref{self2}) determine a self-consistent 
Fermi surface. 
Existence of a Fermi surface breaks the K-invariance inevitably. 
The mean value of the kinetic energy $\langle\varepsilon_l\rangle=
\int_{\rm BZ}\varepsilon_l({\bf p})d^2p/(2\pi)^2$ is equal to 
$-v_l(0)\bar\nu$, which is independent of $r$ and $\theta$. 
The energy per particle in the $l$-th Landau level is calculated as 
\begin{equation}
E^ {(l)}={1\over2\bar\nu}\sum_{\rm X}\vert U_l({\bf X})
\vert^2\{{\tilde v}_l(2\pi\hat{\bf X})-v_l(\hat{\bf X})\}.
\label{ef}
\end{equation}
$E^{(l)}$ is a function of $\bar\nu$, $r$ and $\theta$. 
The parameters $r$ and $\theta$ are determined so as to 
minimize the energy $E^{(l)}$ at a fixed $\bar\nu$. 

\begin{figure}[h]
\centerline{
\epsfysize=5in\epsffile{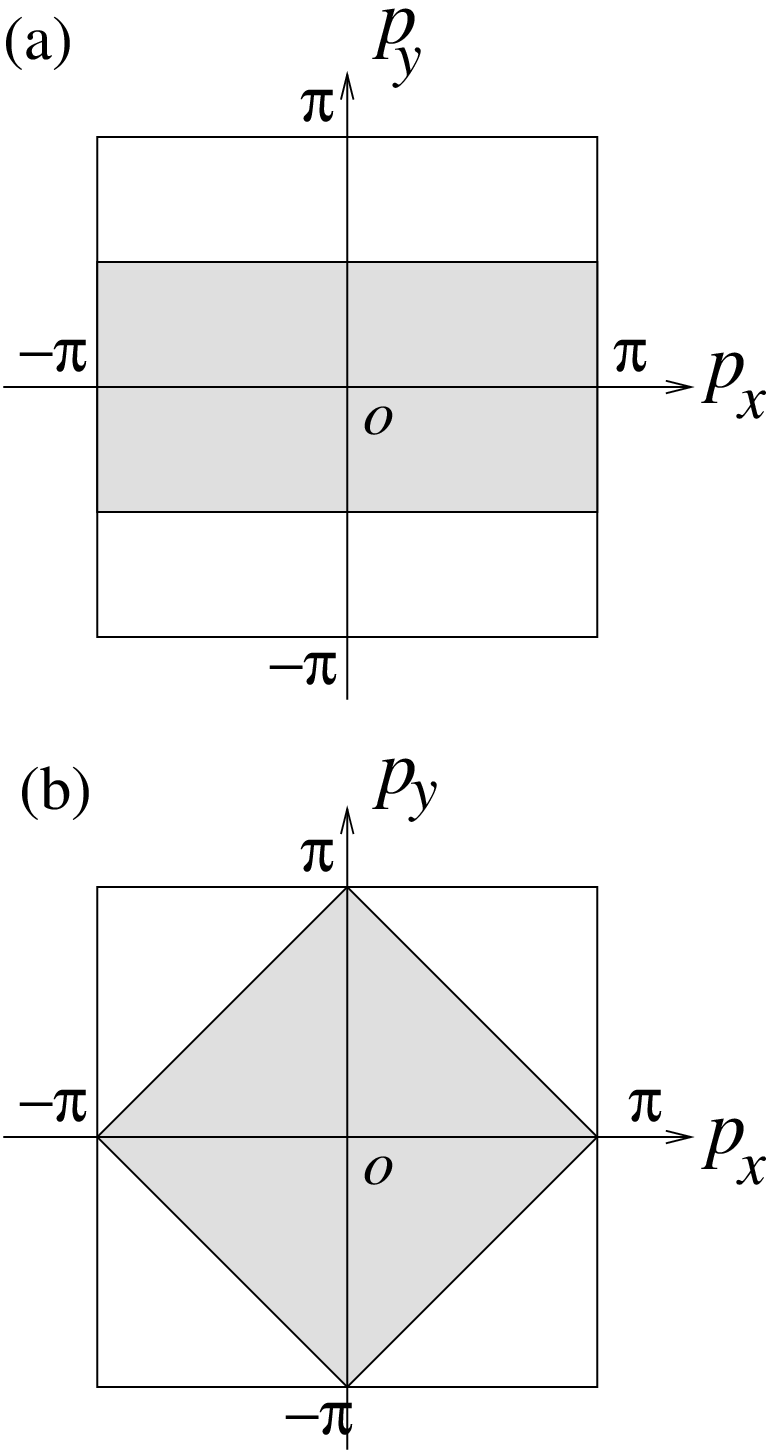}}
FIG. 2 The belt-shaped Fermi-sea (shaded region) 
for the UCDW at the half-filling. 
(b) The diamond-shaped Fermi-sea (shaded region) 
for the CWC at the half-filling. 
\end{figure}
There are two types of the Fermi-sea which satisfies the 
Eqs.~(\ref{self1}) and (\ref{self2}). 
The one is a belt-shaped Fermi-sea illustrated in Fig.~2 (a), 
that is, $\vert p_y\vert\leq p_{\rm F}$. 
This solution corresponds to the UCDW state. 
At $\nu=l+\bar\nu$, the Fermi wave number $p_{\rm F}$ is 
equal to $\pi\bar\nu$ and the mean field of the UCDW becomes 
\begin{equation}
U_l({\bf X})=\delta_{m,0}{\sin(p_{\rm F}n)\over \pi n},
\label{sin}
\end{equation}
for ${\bf X}=(m,n)$. 
We take $\theta=\pi/2$ for the UCDW without losing 
generality because the system has the rotational invariance. 
Then the charge density of the UCDW 
is uniform in the y-direction and oscillates in the x-direction 
with a wave length $\lambda_{\rm CDW}=ra$. 
In the y-direction, $p_{\rm F}$ corresponds to the Fermi wave 
number $k_{\rm F}=p_{\rm F}r/a=\pi{\bar\nu}r/a$ in the real space. 
The duality relation between $\lambda_{\rm F}=2\pi/k_{\rm F}$ 
and $\lambda_{\rm CDW}=ra$ exists. 
Namely $\lambda_{\rm F}\lambda_{\rm CDW}=2a^2/\bar\nu$. 
In the CF model, the composite fermion has a wave number 
$k_{\rm F}=\sqrt{2\pi}/a$ at $\nu=1/2$. 
In the UCDW state for $r=1.636$ which minimizes the energy at 
$\nu=1/2$,\cite{x} $\pi{\bar\nu}r$ 
equals 2.57, which is very close to the value $\sqrt{2\pi}=
2.51$. 
This implies that there exists unknown connection between the 
CF state and the UCDW state. 

The other Fermi-sea which satisfies the 
Eqs.~(\ref{self1}) and (\ref{self2}) at $\bar\nu=1/2$ is a 
diamond-shaped one illustrated in Fig.~2 (b). 
This solution corresponds to the CWC state whose density is 
modulated with the same periodicity as the von Neumann lattice. 
The mean field of the CWC becomes 
\begin{equation}
U_l({\bf X})={2\over(\pi)^2}{\sin{\pi\over2}(m+n)\sin{\pi\over2}(m-n)
\over m^2-n^2},
\label{sisi}
\end{equation}
for ${\bf X}=(m,n)$. 
Substituting Eqs.~(\ref{sin}) and (\ref{sisi}) into Eq.~(\ref{ef}), 
we calculate the energy for various CCDW states at ${\bar\nu}=1/2$. 
By varying the parameters $r$ and $\theta$, we obtained the lowest 
energy numerically. 

The results are summarized in Table.~I. 
The unit of the energy is $q^2/l_B$ and $l_B=\sqrt{1/eB}$. 
As seen in the Table, the UCDW state is the lowest energy state 
in all cases. 
Therefore the UCDW state is the most plausible state in the 
CCDW states. 
For the CWC state, $\theta=\pi/2$ corresponds to a rectangular 
lattice and $\theta=\pi/3$ to a triangular lattice. 
($\theta=\pi/2$ and $r=1$ means a square lattice and 
$\theta=\pi/3$ and $r=1.075$ means a regular triangular lattice.) 
For the UCDW state, the wave length $\lambda_{\rm CDW}=ra$ 
increases with the increasing $l$. 
This behavior is consistent with the numarical 
calculation in finite systems.\cite{paa} 
The Hartree-Fock calculation in the higher Landau level\cite{o,p} 
predicts $\lambda_{\rm CDW}=2.7\sim2.9 \sqrt{2l+1}l_B$. 
Our results agree with this at $l=1,2,3$. 
At $l=0$, however, our result is much smaller than this. 

At $\nu=1/2$, the energy of the UCDW state is slightly higher 
than the value of the gapfull charge density wave (CDW) 
calculation.\cite{y} 
In the gapfull CDW state, the higher order correction is 
small because of the energy gap. 
In the UCDW state, however, the correction might be large 
compared with the CDW state. 
Therefore it is necessary for more definite conclusion to 
include fluctuation effects around the mean field. 
Although this task goes beyond the scope of this paper, 
the Hamiltonian on the von Neumann lattice must be useful 
to study fluctuation effects. 

\section{Property of the UCDW states}
In this section, we calculate the density profile of the 
UCDW and wave length dependence of the energy. 
The density of the electron for $\nu=l+1/2$ reads
\begin{equation}
\rho({\bf r})=l+\int_{-\pi}^{\pi}{d p_x\over 2\pi}
\int_{-\pi/2}^{\pi/2}{d p_y\over 2\pi}
\vert u_{l,{\bf p}}({\bf r})\vert^2.
\label{rr}
\end{equation}
Since $\vert u_{l,\bf p}({\bf r})\vert^2$ is a periodic function of 
$\bf p$ in the BZ and depends only on the 
combination of ${\bf r} +(p_y/2\pi){\bf e}_1- (p_x/2\pi){\bf e}_2$, 
$\rho$ of Eq.~(\ref{rr}) is uniform in the y-direction. 
Here we take ${\bf e}_1=(r,0)$ and ${\bf e}_2=(0,1/r)$ for 
$\theta=\pi/2$. 
The translation in the momentum space is equivalent to the 
translation of the charge density of the CCDW. 
Therefore the symmetry breaking of the K-invariance is same as 
the symmetry breaking of the translational invariance in the real space. 

\vspace{-3cm}
\begin{figure}[h]
\centerline{
\epsfysize=4.5in\epsffile{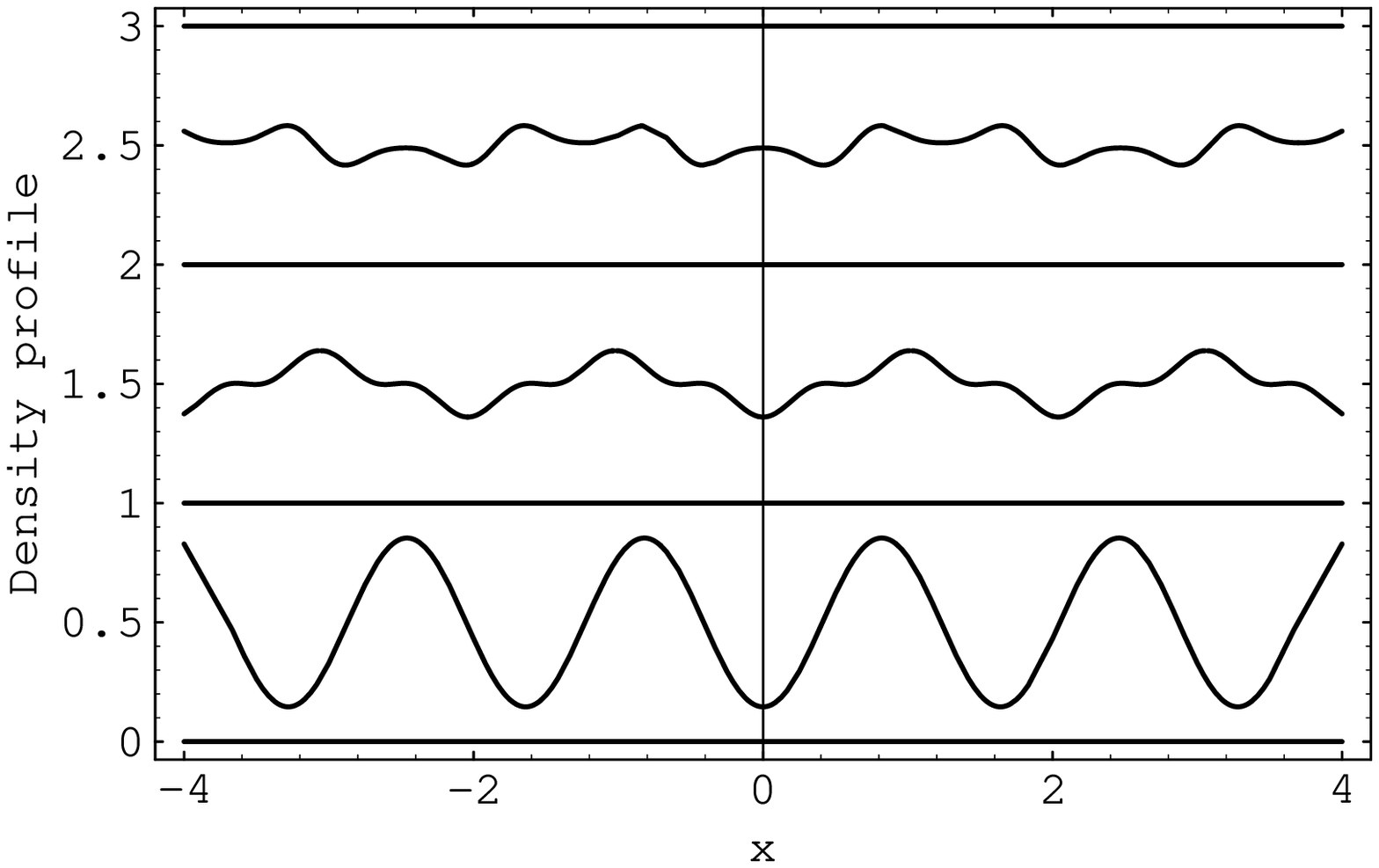}}
\vspace{-3cm}
FIG. 3 The density profile function $\rho(x)$ for $\nu=$0, 0.5, 1, 
1.5, 2, 2.5, and 3. The units of $\rho$ and $x$ are $a^{-2}$ and $a$ 
respectively.
\end{figure}
The density profiles for the UCDW at the half-filled $l$-th Landau 
level of $l=0,1,2$ are plotted in Fig.~3. 
The unit of the density is $a^{-2}$ and the wave length $\lambda_{
\rm CDW}=ra$ in Table.~I is used in Fig.~3. 
As seen in this Figure, the amplitude of the wave decreases 
with the increasing $l$. 

\vspace{-3cm}
\begin{figure}[h]
\centerline{
\epsfysize=4.5in\epsffile{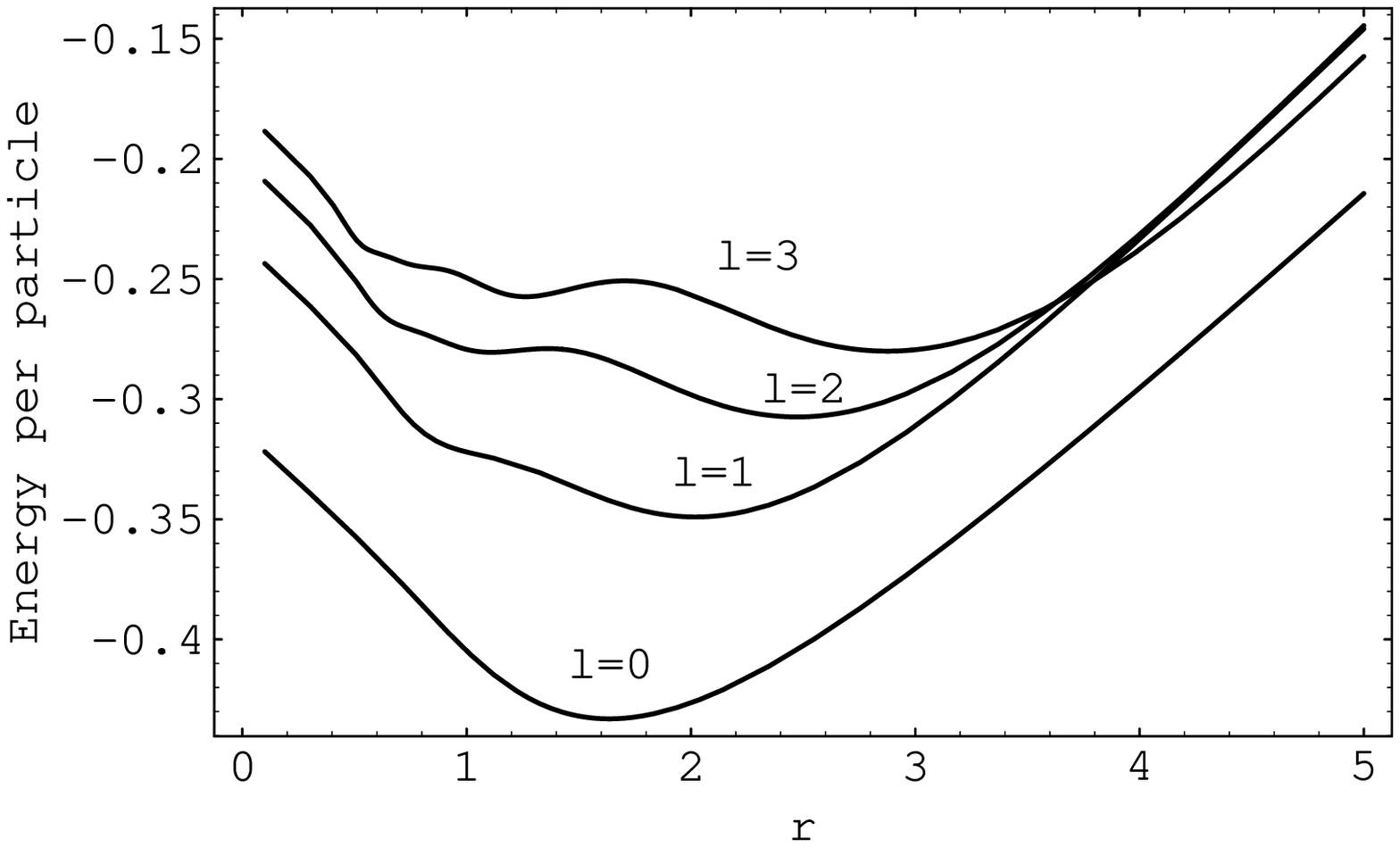}}
\vspace{-3cm}
FIG. 4 The energy per particle $E^{(l)}$ 
as a function of $r$ for $\nu=l+1/2$.   
The units of the energy is $q^2/l_B$. 
\end{figure}
\begin{tabular}{cccc}\hline\hline 
$\qquad l\qquad$ &\  $E_{\rm UCDW}$\  &\  $E_{\rm CWC}(\pi/2)$\ & 
\ $E_{\rm CWC}(\pi/3)\ $ \\
\hline 
0 & -0.4331 & -0.3939 & -0.3891 \\
1 & -0.3490 & -0.3122 & -0.3110 \\
2 & -0.3074 & -0.2715 & -0.2703 \\
3 & -0.2800 & -0.2448 & -0.2436 \\
\hline\hline
$\qquad l\qquad$ &\ $r_{\rm UCDW}$\ & 
\ $r_{\rm CWC}(\pi/2)\ $ & 
\ $r_{\rm CWC}(\pi/3)\ $ \\
\hline 
0 & 1.636 & 1.000 & 1.295 \\
1 & 2.021 & 1.000 & 1.075 \\
2 & 2.474 & 1.000 & 1.075 \\
3 & 2.875 & 1.205 & 1.335 \\
\hline\hline
\end{tabular}

\noindent
Table I. Minimum energy and corresponding parameter $r$ for the 
CCDW states at $\nu=l+1/2$. 
For the CWC states, the energies at $\theta=\pi/2$ 
(rectangular lattice) and $\pi/3$ (triangular lattice) are shown. 
The units of the energy and $r$ are $q^2/l_B$ and 
$a$ respectively.

\ 

To minimize the energy, we calculated the $r$ dependence of the 
energy which is plotted in Fig.~4. 
As seen in Fig.~4, a quasi-stable state appears for $l\geq 2$ near 
$r=1$ and the $r$ dependence of the energy becomes flattened as 
$l$ grows. 
This means that the UCDW state for $l\geq 2$ has flexibility 
against disorder effect. 
This observation agrees with the absence of the spontaneous 
formation of the anisotropic state for $l\leq1$. 

The kinetic energy $\varepsilon_l({\bf p},{\bar\nu})$ for the 
UCDW is written as
\begin{equation}
\varepsilon_l({\bf p},{\bar\nu})=\int^{\pi}_{-\pi}
{dp'_x\over 2\pi}
\int^{\pi\bar\nu}_{-\pi\bar\nu}{dp'_y\over 2\pi}{\tilde v}_l^{
\rm HF}({\bf p}'-{\bf p}).
\end{equation}
This is independent of $p_x$ after integration and we 
denote $\varepsilon_l({\bf p},{\bar\nu})=\varepsilon_l(p_y,
{\bar\nu})$. 
$\varepsilon_l(p_y,1/2)$ for $l=$0, 1, and 2 are shown in Fig.~5. 
\vspace{-3cm}
\begin{figure}[h]
\centerline{
\epsfysize=4.5in\epsffile{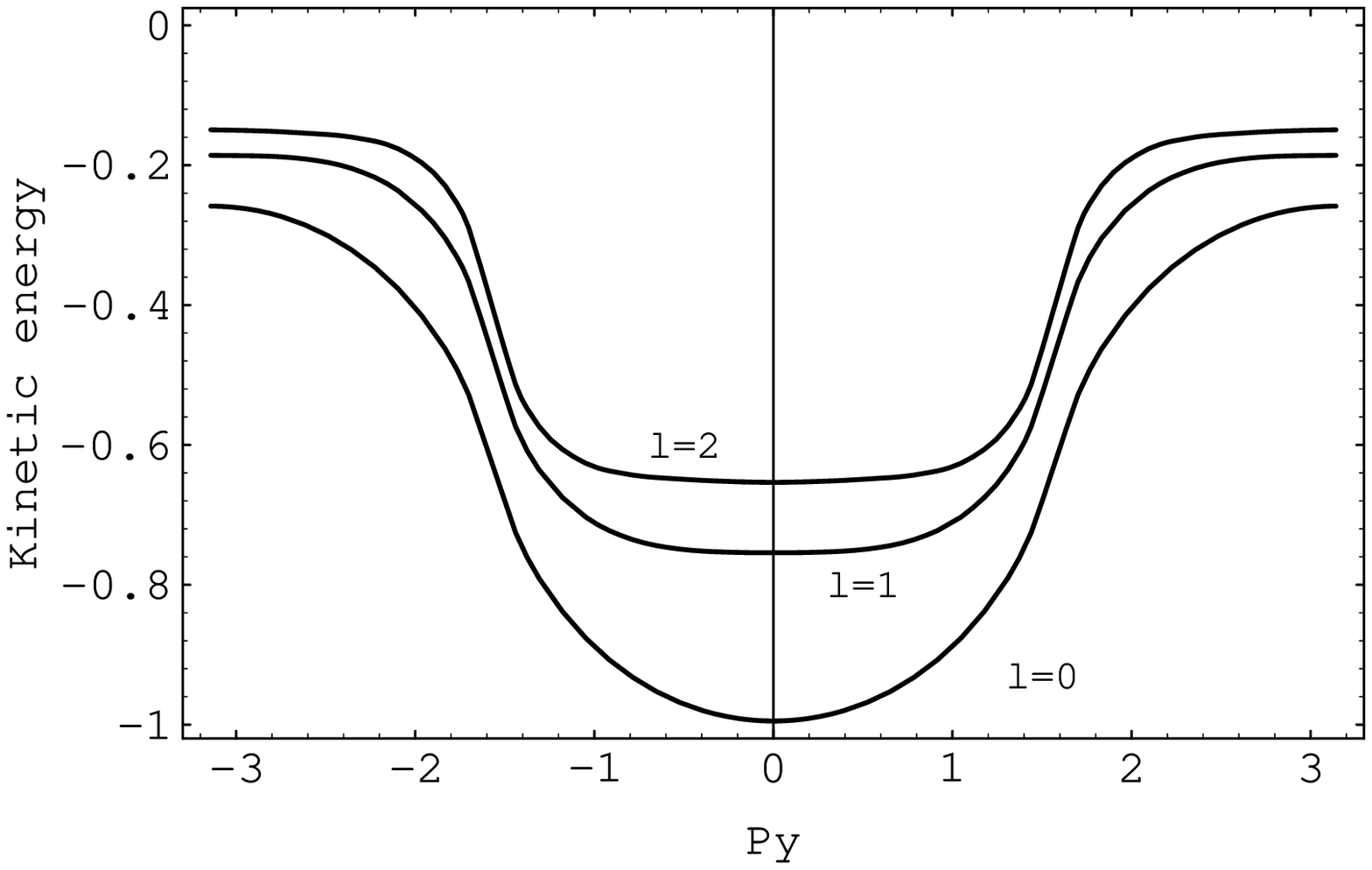}}
\vspace{-3cm}
FIG. 5 The kinetic energy $\varepsilon_l(p_y,1/2)$ for $l=$0, 
1,and 2. 
The units of the energy is $q^2/l_B$. 
\end{figure}
As seen in Fig.~5, the bandwidth $\Gamma_l$ decreases with the 
increasing $l$, that is, 
$\Gamma_0=0.7363$, 
$\Gamma_1=0.5682$, and 
$\Gamma_2=0.5042$ 
in the unit of $q^2/l_B$. 
Using the mean field of Eq.~(\ref{sin}), the kinetic term in 
$H^{(l)}_{\rm HF}$ is written as 
\begin{equation}
K^{(l)}_{\rm HF}=\sum_m\int{dp_y\over 2\pi} 
a^\dagger_{l,m}(p_y)\varepsilon_l(p_y,\bar\nu)a_{l,m}(p_y),
\end{equation}
where $a_{l,m}(p_y)=\sum_n b_l({\bf X})e^{-ip_y n}$ for ${\bf X}=(m,n)$. 
Therefore the UCDW state is regarded as a collection of the 
one-dimensional lattice Fermi-gas systems which extend to the 
y-direction. 
In the Buttiker-Landauer formula,\cite{BL} the conductance of a 
one-dimensional channel is equal to $e^2/2\pi$ in the absence of 
the backscattering effect. 
Thus the conductance of the UCDW have an anisotropic value as
\begin{eqnarray}
\sigma_{xx}&=&0,\\
\sigma_{yy}&=&n_x {e^2\over 2\pi},
\end{eqnarray}
where $n_x$ is a number of the one-dimensional channels which extend 
from one edge to the other edge. 
If we take $\sigma_{xy}=\nu e^2/2\pi$, the resistance becomes 
\begin{eqnarray}
\rho_{xx}&=&{n_x\over\nu^2}{2\pi\over e^2},\\
\rho_{yy}&=&0.
\end{eqnarray}
Thus the formation of the UCDW leads the anisotropy in the 
magnetoresistance. 
For $\nu=9/2$, $2\pi/e^2\nu^2=1.3\times10^3$ $\Omega$ which 
is of the same order as the experimental value $\sim1.0\times10^3$ 
$\Omega$. 
Disorder effects decreases $n_x$ by destroying the UCDW 
ordering. 
Furthermore the backscattering effect due to impurities 
reduces $\sigma_{yy}$ and $\rho_{xx}$. 
In the case of the edge modes in the quantum Hall regime, 
there is no backscattering because of the chirality. 
The left-mover lives on the one edge far from the other 
edge where the right-mover lives. 
For the UCDW state, on the other hand, each one-dimensional 
system has the width of $ra$ at most. 
Therefore the backscattering effect strongly affects on the 
conductance in the UCDW. 

To conclude this section, we point out a subtle problem 
concerned with the K-invariance and sliding mode. 
In an ordinary one-dimensional system, the difference of the 
chemical potentials between the left and right edge of the 
Fermi sea yields the net electric current. 
In the one-dimensional system of the UCDW, however, the difference 
of the chemical potentials can be canceled out by sliding the 
Fermi sea in Eqs.~(\ref{self1}) and (\ref{self2}), thanks to the 
K-invariance. 
Then there is no net electric current. 
This contradicts the above assertion apparently. 
As mentioned before, the translation in the momentum space is 
equal to the translation of the charge density in the real space. 
In other words, to slide the Fermi sea is the same as to slide the 
CCDW in the real space. 
The sliding mode is expected to be pinned by impurities. 
Therefore the violation of the K-invariance due to pinning 
of the CCDW can remedy the contradiction. 

\section{Summary and Discussion}

In this paper we have studied the CCDW state whose periodicity of 
the charge density coincides with that of the von Neumann lattice.
The CCDW state is gapless and has an anisotropic Fermi surface. 
We obtained two types of the CCDW state, the UCDW state and CWC state.  
By calculating the Hartree-Fock energy, the UCDW is found to have a lower 
energy at the half-filled Landau levels. 
Furthermore, the wave length dependence of the energy, the 
density profile, and the kinetic energy of the UCDW 
are calculated numerically. 
As a result, it is found that the amplitude of the UCDW 
and bandwidth of the Landau level decrease with 
the increasing $l$ and a quasi-stable state appears for $l\geq2$. 

The UCDW state has a belt-shaped Fermi-sea. 
Consequently the system consists of many one-dimensional lattice 
Fermi-gas systems which extend to the uniform direction. 
Formation of this structure could be the origin of the anisotropy 
observed in experiments. 
To confirm this speculation, experimental works 
for detecting the wave length of the UCDW and theoretical works to 
include fluctuations around the mean field solution are necessary. 
Since there is no energy gap in the CCDW state, the fluctuation 
effect might be large compared with the gapfull CDW state. 
Actually Fradkin and Kivelson proposed a rich phase diagram 
by considering fluctuations around the stripe-ordered state. 
We believe that the von Neumann lattice formalism presents an 
appropriate scheme to study the fluctuation effects around 
the mean field. 

Recently a new insulating state is discovered around the quarter-filled 
third Landau level.\cite{z} 
This state seems gapfull and to have an integral quantized Hall 
conductance. 
These facts suggest that the state is a gapfull CDW 
state which is different from the CCDW. 
The CDW state whose periodicity 
is $q/p$ times that of the von Neumann lattice is gapfull and 
has $q$ bands with $p$-hold degeneracy.\cite{y,v} 
In the presence of a magnetic field and periodic potential, 
the Hall conductance of the free electron system in the gap region 
is equal to the Chern number.\cite{za,v} 
Thus the observed quantized Hall conductance is probably the 
Chern number of the periodic potential problem. 
The transition between this gapfull CDW and CCDW studied in this paper 
is an interesting future problem.

\acknowledgements

I would like to thank K. Ishikawa, T. Ochiai, and 
J. Goryo for useful discussions. 
I also thank Y. S. Wu and G. H. Chen for helpful discussions. 
This work was partially supported by the special Grant-in-Aid for 
Promotion of Education and Science in Hokkaido University provided 
by the Ministry of Education, Science, Sports, and Culture, 
the Grant-in-Aid for Scientific Research on Priority area (Physics of 
CP violation) (Grant No. 11127201), the Grant-in-Aid for 
Basic Research (Grant No. 10044043), and the special Grant for Basic 
Research (Hierarchical matter analyzing system) from the Ministry of 
Education, 
Science, Sports, and Culture, Japan.

\end{multicols}
\end{document}